\documentclass[iop,revtex4,numberedappendix,appendixfloats,twocolappendix]{emulateapj}
\usepackage{times}
\usepackage{appendix}
\usepackage{pdflscape}

\begin{document}

\newcommand{\oii}{\text{[\ion{O}{2}]}}
\newcommand{\neiii}{\text{[\ion{Ne}{3}]}}
\newcommand{\oiii}{\text{[\ion{O}{3}]}}
\newcommand{\woiii}{\text{$W_\lambda(\oiii)$}}
\newcommand{\nii}{\text{[\ion{N}{2}]}}
\newcommand{\hei}{\text{\ion{He}{1}}}
\newcommand{\heii}{\text{\ion{He}{2}}}
\newcommand{\ha}{\text{H$\alpha$}}
\newcommand{\wha}{\text{$W_\lambda(\ha)$}}
\newcommand{\hb}{\text{H$\beta$}}
\newcommand{\hg}{\text{H$\gamma$}}
\newcommand{\hd}{\text{H$\delta$}}
\newcommand{\he}{\text{H$\epsilon$}}
\newcommand{\hz}{\text{H$\zeta$}}
\newcommand{\hn}{\text{H$\eta$}}
\newcommand{\htheta}{\text{H$\theta$}}
\newcommand{\hiota}{\text{H$\iota$}}
\newcommand{\pa}{\text{Pa$\alpha$}}
\newcommand{\pb}{\text{Pa$\beta$}}
\newcommand{\pg}{\text{Pa$\gamma$}}
\newcommand{\pd}{\text{Pa$\delta$}}
\newcommand{\hi}{\text{\ion{H}{1}}}
\newcommand{\hii}{\text{\ion{H}{2}}}
\newcommand{\hk}{\text{H$\kappa$}}
\newcommand{\caii}{\text{\ion{Ca}{2}}}
\newcommand{\sii}{\text{[\ion{S}{2}]}}
\newcommand{\wlya}{\text{$W_\lambda$({\rm Ly$\alpha$})}}
\newcommand{\wlyaem}{\text{$W_\lambda^{\rm em}$({\rm Ly$\alpha$})}}
\newcommand{\llya}{\text{$L$(Ly$\alpha$)}}
\newcommand{\llyaobs}{\text{$L$(Ly$\alpha$)$_{\rm obs}$}}
\newcommand{\llyaint}{\text{$L$(Ly$\alpha$)$_{\rm int}$}}
\newcommand{\lyafrac}{\text{$f_{\rm esc}^{\rm spec}$(Ly$\alpha$)}}
\newcommand{\lha}{\text{$L$(H$\alpha$)}}
\newcommand{\lhb}{\text{$L$(H$\beta$)}}
\newcommand{\sfrha}{\text{SFR(\ha)}}
\newcommand{\sfrsed}{\text{SFR(SED)}}
\newcommand{\ssfrha}{\text{sSFR(\ha)}}
\newcommand{\ssfrsed}{\text{sSFR(SED)}}
\newcommand{\ebmvneb}{E(B-V)_{\rm neb}}
\newcommand{\ebmvcont}{E(B-V)_{\rm cont}}
\newcommand{\ebmvlos}{E(B-V)_{\rm los}}
\newcommand{\nhi}{N(\text{\ion{H}{1}})}
\newcommand{\lognhi}{\log[\nhi/{\rm cm}^{-2}]}
\newcommand{\lognhitable}{\log\left[\frac{\nhi}{{\rm cm}^{-2}}\right]}
\newcommand{\lya}{\text{Ly$\alpha$}}
\newcommand{\lyb}{\text{Ly$\beta$}}
\newcommand{\lyg}{\text{Ly$\gamma$}}
\newcommand{\comment}[1]{}
\newcommand{\wciii}{\text{$W_\lambda$(\ion{C}{3}])}}
\newcommand{\ciii}{\text{\ion{C}{3}]}}
\newcommand{\interoiii}{\text{\ion{O}{3}]}}
\newcommand{\rsiione}{R(\text{\ion{Si}{2}}\lambda 1260)}
\newcommand{\rsiitwo}{R(\text{\ion{Si}{2}}\lambda 1527)}
\newcommand{\siii}{\text{\ion{Si}{2}}}
\newcommand{\cii}{\text{\ion{C}{2}}}
\newcommand{\civ}{\text{\ion{C}{4}}}
\newcommand{\rcii}{R(\text{\ion{C}{2}}\lambda 1334)}
\newcommand{\ralii}{R(\ion{Al}{2}\lambda 1670)}
\newcommand{\qh}{Q(\text{H$^0$})}
\newcommand{\rs}{{\cal R}_{\rm s}}
\newcommand{\fcovhi}{f_{\rm cov}(\hi)}
\newcommand{\fcovmetal}{f_{\rm cov}({\rm metal})}
\newcommand{\fesclya}{f_{\rm esc}^{\rm spec}(\lya)}
\newcommand{\logxi}{\log[\xi_{\rm ion}/{\rm s^{-1}/erg\,s^{-1}\,Hz^{-1}}]}
\newcommand{\lir}{L_{\rm IR}}
\newcommand{\lbol}{L_{\rm bol}}
\newcommand{\luv}{L_{\rm UV}}

\title{\bf P\lowercase{aschen-line} C\lowercase{onstraints on} D\lowercase{ust} A\lowercase{ttenuation and} S\lowercase{tar} F\lowercase{ormation at} \lowercase{$z\sim 1-3$} \lowercase{with} {\bf \em JWST}/NIRS\lowercase{pec}}

%{\rm JWST}/NIRS\lowercase{pec} P\lowercase{aschen-line} C\lowercase{onstraints on} D\lowercase{ust} A\lowercase{ttenuation and} S\lowercase{tar} F\lowercase{ormation in} H\lowercase{igh-redshift} G\lowercase{alaxies}}

\author{\sc Naveen A. Reddy\altaffilmark{1},
Michael W. Topping\altaffilmark{2},
Ryan L. Sanders \altaffilmark{3,4},
Alice E. Shapley \altaffilmark{5},
and Gabriel Brammer \altaffilmark{6,7}}

\altaffiltext{1}{Department of Physics and Astronomy, University of California, Riverside, 900 University Avenue, Riverside, CA 92521, USA; naveenr@ucr.edu}
\altaffiltext{2}{Steward Observatory, University of Arizona, 933 North Cherry Avenue, Tucson, AZ 85721, USA}
\altaffiltext{3}{Department of Physics, University of California, Davis, One Shields Ave, Davis, CA 95616, USA}
\altaffiltext{4}{NASA Hubble Fellow}
\altaffiltext{5}{Department of Physics and Astronomy, University of California, Los Angeles, 430 Portola Plaza, Los Angeles, CA 90095, USA}
\altaffiltext{6}{Cosmic Dawn Center (DAWN)}
\altaffiltext{7}{Niels Bohr Institute, University of Copenhagen, Jagtvej 128, 2200 Copenhagen N, Denmark}

\slugcomment{DRAFT: \today}

\begin{abstract}

We use medium resolution {\em JWST}/NIRSpec observations from the
Cosmic Evolution Early Release Science (CEERS) Survey to place the
first constraints on dust attenuation and star formation based on the
Paschen lines for a sizable sample of 63 galaxies at redshifts
$z=1.0-3.1$.  Our analysis indicates strong correlations between the
Balmer decrement, $\ha/\hb$, and line ratios that include the Paschen
lines (i.e., $\pa/\hb$, $\pb/\hb$, and the Paschen decrement,
$\pa/\pb$), suggesting that the former is sensitive to the overall
dust obscuration towards $\hii$ regions in high-redshift galaxies.
The line ratios are used to derive the nebular reddening, $\ebmvneb$,
and star-formation rates (SFRs).  There is marginal evidence that SFRs
deduced from the Paschen lines may exceed by $\approx 25\%$ those
derived from the Balmer lines alone, suggesting the presence of star
formation that is optically thick in the Balmer lines, though deeper
observations are needed to confirm this result.  Using the
Paschen-line constraints on bolometric SFRs, we reevaluate the
relationship between dust obscuration and UV spectral slope, and find
a reddening of the UV continuum that, on average, follows the SMC
extinction curve.  This analysis highlights the need for deeper
spectroscopy of more representative samples to evaluate nebular dust
attenuation and bolometric SFRs in high-redshift galaxies, and their
relationship to the reddening of the UV continuum.

\end{abstract}


\section{\bf INTRODUCTION}
\label{sec:intro}

Hydrogen recombination lines provide critical constraints on the
star-formation rates (SFRs) and dust reddening towards the ionized
($\hii$) regions of high-redshift star-forming galaxies
\citep{kennicutt94}. These lines are dependent on the presence of
short-lived massive stars and are therefore less sensitive to
star-formation history (SFH) than the UV continuum.  The lines suffer
less extinction than the far-UV continuum, and can be measured from
the ground with high signal-to-noise ($S/N$) for individual typical
star-forming galaxies up to redshift $z\sim 2.6$.  Constraints on the
reddening of $\hii$ regions (i.e., nebular reddening) are fundamental
to the interpretation of some rest-frame optical emission lines ratios
that are commonly used to infer the ionization state of the ISM and
gas-phase metallicity (e.g., $\oiii\,\lambda 5008/\oii\,\lambda\lambda
3727,3730$, and R23, defined as $(\oii\,\lambda\lambda 3727,3730 +
\oiii\,\lambda\lambda 4960, 5008)/\hb$).  Moreover, the nebular
reddening and its connection to the reddening of the stellar continuum
can yield insights on the spatial distribution of dust relative to stars.

Observations from ground-based multi-object near-IR spectrographs
(e.g., Keck/MOSFIRE, VLT/KMOS) and space-based facilities (e.g.,
HST/WFC3 grism) have yielded measurements of hydrogen recombination
lines and nebular reddening for thousands of galaxies up to $z\sim 2.6$
(e.g., \citealt{forster09, kashino13, price14, reddy15, debarros16,
  reddy20, shivaei20a, rezaee21, battisti22}).  There is a general
consensus from these studies that, on average, the nebular reddening
exceeds that of the stellar continuum, where the difference in
reddening may correlate with SFR, specific SFR (sSFR), stellar mass,
and/or gas-phase metallicity.  These differences in the nebular and
stellar continuum reddening are typically attributed to the higher
dust column densities expected for the lines of sight towards young
massive stars which are embedded in their parent molecular clouds
(e.g., \citealt{calzetti94}).

As useful as these spectroscopic observations have been for advancing
our understanding of the nebular dust attenuation and
recombination-line SFRs of high-redshift galaxies, virtually all of
this knowledge is based on the lowest-order lines of the Balmer
series, in particular $\ha$ and $\hb$, and only for galaxies at
redshifts $z\la 2.6$.  Up until recently, Paschen lines have been
robustly detected in just a few $z>2$ lensed galaxies
\citep{papovich11, finkelsteink11}.  The recently-launched {\em James
  Webb Space Telescope} ({\em JWST}) and its NIRSpec instrument
\citep{ferruit22} now offer the wavelength coverage and sensitivity to
detect higher-order Paschen lines for sub-$L^\ast$ galaxies ($\pa$ at
$\lambda = 1.88$\,$\mu$m up to $z\sim 1.8$, $\pb$ at $\lambda =
1.28$\,$\mu$m up to $z\sim 3.1$), as well as Balmer lines into the
epoch of reionization ($\ha$ up to $z\sim 6.5$, $\hb$ up to $z\sim
9.3$; \citealt{shapley23}).  As they are less affected by dust
attenuation, the Paschen lines more closely track the bolometric SFR
than the (observed) Balmer lines; they may reveal the presence of star
formation that is optically thick in the Balmer lines; and, in
combination with Balmer lines, they can provide unprecedented
constraints on the nebular reddening and characteristics of the ISM
that depend on it (e.g., ionization parameter and gas-phase
metallicity).

In this paper, we use the first spectroscopic data from the Cosmic
Evolution Early Release Science (CEERS) survey \citep{finkelstein22}
with {\em JWST} (Section~\ref{sec:data}) to investigate the improvements in
nebular reddening constraints afforded by the Paschen lines
(Section~\ref{sec:reddening}), and the implications of such
constraints on the reddening of the UV continuum
for galaxies at redshifts $z=1.0-3.1$ (Section~\ref{sec:irx}) .  A
\citet{chabrier03} initial mass function (IMF) is considered
throughout the paper.  Wavelengths are reported in the vacuum frame.
%Magnitudes
%are on the AB system \citep{oke83}.  
We adopt a cosmology with $H_{0}=70$\,km\,s$^{-1}$\,Mpc$^{-1}$,
$\Omega_{\Lambda}=0.7$, and $\Omega_{\rm m}=0.3$.

\section{\bf DATA}
\label{sec:data}

\subsection{Observations}

The publicly available NIRSpec Multi Shutter Assembly (MSA) data from
the CEERS program (Program ID: 1345; \citealt{finkelstein22}) were
used for this analysis.  These medium resolution $R\sim 1000$ data
were obtained using the G140M/F100LP, G235M/F170LP, and G395M/F290LP
grating and filter combinations, covering a wavelength range of
$\lambda = 1-5$\,$\mu$m.  Observations were obtained over six
pointings in the AEGIS field, with each pointing consisting of three
exposures of 14 groups with the NRSIRS2 readout mode, and a total of
3107\,s integration in each grating and filter combination.  Each
``slit'' on the MSA was formed from three adjacent microshutters, and
a three-point nod dither pattern was adopted.  Together, the six
pointings include 321 slits and 318 unique targets.

\subsection{Data Reduction}

NIRSpec spectroscopic data were reduced following the procedures given
in Topping et al. (2023, in prep).  Briefly, individual uncalibrated
exposures were processed through the {\em JWST}
\texttt{calwebb\_detector1}
pipeline\footnote{\url{https://jwst-pipeline.readthedocs.io/en/latest/index.html}}.
This step masks all saturated pixels, subtracts the bias and dark
current, and masks `snowballs' and `showers' resulting from
high-energy cosmic ray hits.  The images were then corrected for
striping by subtracting the $1/f$ noise.  Following these steps, 2D
spectrograms were cut out for each slit on the MSA.  Flat-field
corrections, subtraction of background based on the dithered
exposures, and photometric calibrations were applied to the 2D
spectrograms, and a wavelength solution using the most recent
calibration reference data system (CRDS; \texttt{jwst\_1027.pmap}) was
implemented.  Each spectrogram was then rectified and interpolated
onto a common wavelength grid for its grating and filter combination.
Finally, the rectified and interpolated spectrograms were combined
using the offsets appropriate for the three-point nod dither pattern,
excluding pixels masked during the calibration steps.  The
corresponding 2D error spectra were generated by summing in quadrature
the Poisson noise, read noise, and variance between exposures.
One-dimensional spectra were optimally extracted from the 2D
spectrograms and, in cases where neither emission lines nor continua
were seen, a blind extraction was performed based on the position of
the object determined from other gratings (see Sanders et~al., 2023,
in prep, for further details).  If there were no emission lines or
continua visible in the 2D spectrograms for any grating, then no
extraction was performed.  One-dimensional spectra were extracted for
252 targets.

\subsection{Slit Loss Corrections and Flux Scaling}

The combination of a small NIRSpec MSA microshutter width
($0\farcs20$), off-center location of the target within the
microshutter, and a wavelength-dependent PSF necessitates accurate
corrections for light falling outside the microshutter and 1D
extraction window.  These ``slit losses'' were computed as follows.
If a CEERS target had {\em JWST}/NIRCam F115W imaging, a $12\arcsec \times
12\arcsec$ subimage centered on the target was extracted.  The
corresponding segmentation map was used to mask all pixels belonging
to unrelated objects within the subimage.  The subimage was then
rotated, accounting for both the $45^\circ$ tilt of the AEGIS F115W
mosaic relative to North and the position angle of the NIRSpec MSA
observations.  This rotated subimage was then convolved with Gaussian
kernels to produce the expected light profiles at longer wavelengths.
The full widths at half maximum (FWHMs) of these kernels were computed
by subtracting in quadrature the FWHM of the {\em JWST} PSF at
$1.15$\,$\mu$m from the FWHMs of the {\em JWST} PSFs at $\lambda = 1.2
- 5.3$\,$\mu$m, in increments of $\delta\lambda = 0.1$\,$\mu$m.

If a target did not have F115W imaging, a S\'ersic fit to the
HST/WFC3 F160W image was used to create a model of the galaxy.  The
parameters of the S\'ersic fit were obtained from the GALFIT modeling
performed by \citet{vanderwel14}, and a model was created only if the
fit was deemed acceptable based on the criteria given in that study.
This model was then rotated as described above and convolved with the
{\em JWST} PSFs at $\lambda = 0.6 - 5.3$\,$\mu$m, in increments of
$\delta\lambda = 0.1$\,$\mu$m, as determined from the {\em JWST}
\texttt{WebbPSF}
software\footnote{\url{https://webbpsf.readthedocs.io/en/stable/installation.html}}.
Finally, if a target did not have F115W imaging or an acceptable
S\'ersic model fit to its F160W light profile, a point source was
assumed.

The convolved images of the target at each wavelength were then
shifted according to the expected position of the target within the
microshutter and masked.  The masking accounts for the $0\farcs20$
microshutter width, the $0\farcs07$ gap between adjacent microshutters
along the cross-dispersion axis, and the size of the window used to
extract the 2D spectrograms to 1D.  The 1D spectra were then divided
by the fraction of light passing through the unmasked
region(s)---i.e., the slit and extraction apertures---as a function of
wavelength.

To test the validity of this procedure, we first checked if the slit
loss corrections based on the F115W imaging were consistent with those
obtained by assuming a S\'ersic model of the F160W light profile for
galaxies where both F115W imaging and a F160W-based S\'ersic model
were available.  This comparison shows that the fractions of
transmitted flux obtained with the two methods agree within $20\%$.
In addition, the continuum is significantly detected in the individual
NIRSpec spectra for the vast majority of targets, with a typical $S/N$
per wavelength pixel of $\simeq 5$.  Passing the slit-loss-corrected
1D spectra through the NIRCam broadband photometric filters yields
integrated fluxes that are in excellent agreement with the measured
ground-based near-IR, NIRCam, and/or {\em Spitzer}/IRAC
photometry, with a factor of $\simeq 2$ dispersion.  This agreement
suggests that the relative flux calibration between different gratings
is generally robust, and that ratios of lines observed in different
gratings can be trusted.  All spectra were then scaled such that their
integrated fluxes in the relevant bandpasses match the ground-based
near-IR, NIRCam, and/or IRAC photometry (Sanders et~al., 2023, in
prep).

\subsection{Line Measurements and Composite Spectra}
\label{sec:linemeas}

Emission line fluxes were measured by fitting their profiles with
Gaussian functions on top of a continuum model determined by the
best-fit SED to the broadband photometry (Section~\ref{sec:sed}), as
described in Sanders et~al. (2023, in prep).  In the case of blended
complexes, such as $\nii$\,$\lambda 6550$+$\ha$+$\nii$\,$\lambda
6585$, multiple Gaussians were fit simultaneously, along with a
continuum model determined by the best-fit SED.  This method accounts
for stellar absorption under the HI emission lines.  Uncertainties in
line fluxes were estimated by measuring the spread in fluxes obtained
in many realizations of the spectra, where the science spectrum was
perturbed in each realization by the corresponding error spectrum.

Part of the analysis described here takes advantage of the high $S/N$
afforded by composite spectra.  Composite spectra were generated by
shifting each object's spectrum to the rest frame, converting from
flux density to luminosity density, interpolating to a common
wavelength grid, and averaging luminosity densities at each wavelength
point using $3\sigma$ clipping.  Average line luminosities, line
ratios, and their corresponding uncertainties including sample
variance were calculated by generating many realizations of the
composite spectra.  In each realization, the spectra of individual
objects were perturbed according to their measurement errors, and
composite spectra were constructed by random selection of these
perturbed spectra with replacement.  Line flux measurements on the
composite spectra were performed in a manner similar to that of
individual objects.

\subsection{SED Fitting}
\label{sec:sed}

Stellar population modeling was used to account for stellar absorption
under the hydrogen emission lines (Section~\ref{sec:linemeas}), and to
estimate stellar masses ($M_\ast$).  Ground- and space-based
photometry for each target were drawn from existing multi-wavelength
catalogs in the AEGIS field, primarily from the catalogs produced by
the 3D-HST team \citep{momcheva16, skelton14} and those produced by
G. Brammer\footnote{\url{https://s3.amazonaws.com/grizli-v2/JwstMosaics/v4/index.html}}.
The photometry was corrected for the contribution from strong emission
lines.  The modeling employed the FAST program \citep{kriek09} with
the stellar population synthesis models of \citet{conroy09}.
Motivated by previous results (e.g., \citealt{reddy18a, shivaei20a}),
the photometry was modeled with two sets of assumptions: the SMC
extinction curve \citep{gordon03} and a subsolar stellar metallicity
($Z_\ast = 0.004$; ``SMC+subsolar'' modeling), and the
\citet{calzetti00} attenuation curve and a $\sim$solar metallicity
($Z_\ast = 0.02$; ``Calzetti+solar'' modeling).  Stellar masses
derived with the SMC+subsolar modeling are on average $\approx 12\%$
larger than those derived with the Calzetti+solar modeling.  The
amount of stellar absorption under the hydrogen emission lines changes
by only a few percent between the two modeling assumptions.
Therefore, stellar masses and the stellar absorption (which are the
salient parameters in this analysis) are not significantly affected by
the two modeling assumptions.  In Sections~\ref{sec:sfrnebcompare} and
\ref{sec:irx}, we discuss the stellar-metallicity dependence of the
factors used to convert $\ha$ and UV luminosity to SFR.
%Following
%\citet{du18}, we adopted the Calzetti+solar modeling at $z\le 1.4$.
%At $1.4<z\le 3.1$, we adopted the Calzetti+solar modeling for galaxies
%with $\log[M_\ast/M_\odot] > 10.45$ and the SMC+subsolar modeling for
%galaxies at lower masses, where the mass limit is computed from the
%Calzetti+solar modeling.  
Delayed-$\tau$ star-formation histories,
where ${\rm SFR(t)} \propto t \times \exp(-t/\tau)$, were assumed.
Here, $t$ is the time since the onset of star formation and $\tau$ is
the characteristic star formation timescale.  Further details on the
SED modeling are provided in \citet{shapley23}.

\begin{figure}
  \epsscale{1.0}
  \plotone{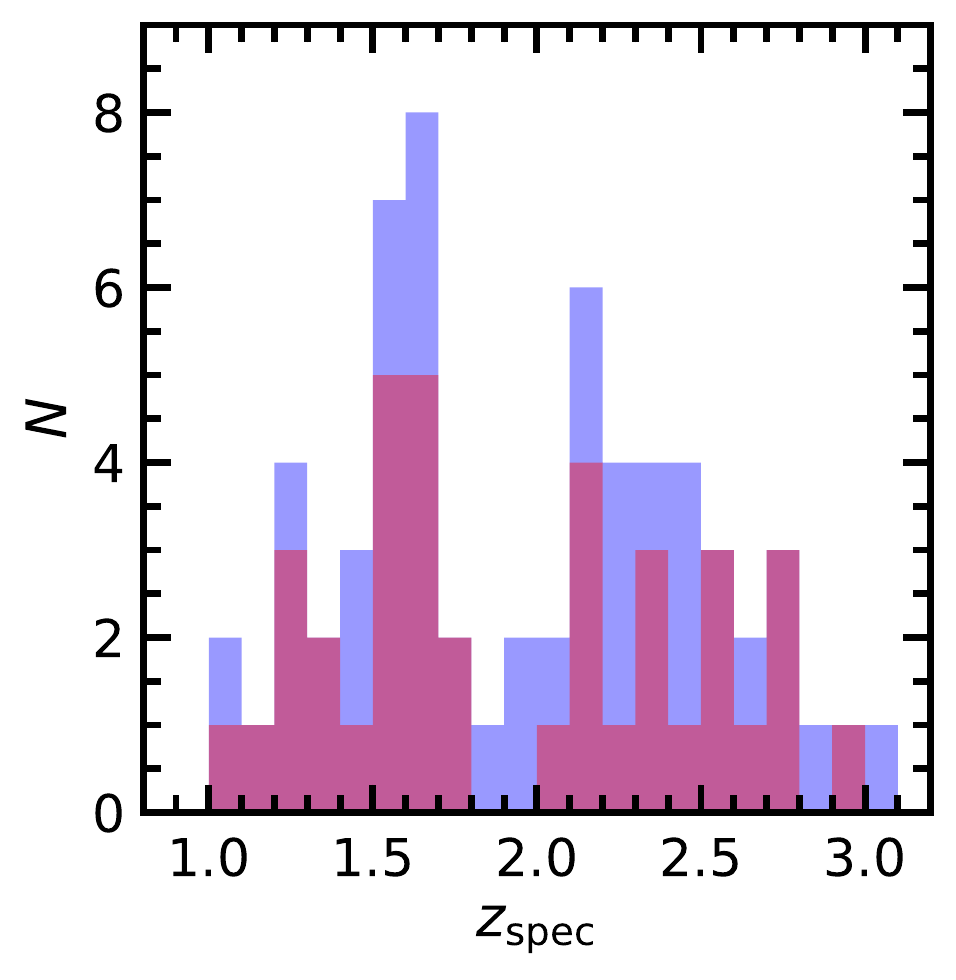}
    \caption{Redshift histogram of the 63 galaxies at $z=1.0-3.1$
      considered in this analysis (blue shaded histogram).  The red
      histogram indicates the 38 objects with $\ge 3\sigma$ detections
      of $\ha$, $\hb$, and either $\pa$ or $\pb$.}
    \label{fig:zhist}
\end{figure}

\begin{deluxetable}{lcccc}
\tabletypesize{\footnotesize}
\tablewidth{0pc}
\tablecaption{Balmer and Paschen Recombination Lines}
\tablehead{
\colhead{Line} &
\colhead{} &
\colhead{$\lambda$ (\AA)\tablenotemark{a}} &
\colhead{} &
\colhead{$I$\tablenotemark{b}}}
\startdata
\ha & & 6564.60 & & 2.790 \\
\hb & & 4862.71 & & 1.000 \\
\hg & & 4341.69 & & 0.473 \\
\hd & & 4102.89 & & 0.263 \\
\pa & & 18756.4 & & 0.305 \\
\pb & & 12821.6 & & 0.155 \\
\pg & & 10941.2 & & 0.087 \\
\pd & & 10052.6 & & 0.054
\enddata
\tablenotetext{a}{Rest-frame vacuum wavelength, taken from the Atomic Spectra Database website of
the National Institute of Standards and Technology (NIST), https://www.nist.gov/pml/atomic-spectra-database.}
\tablenotetext{b}{Intensity of line relative to $\hb$ for Case B recombination, $T_{\rm e}=10,000$\,K,
and $n_{\rm e}=100$\,cm$^{-3}$, based on photoionization modeling with CLOUDY version 17.02 \citep{ferland17}.}
\label{tab:nebinfo}
\end{deluxetable}

\subsection{Final Sample}

Several selection criteria were used to assemble the final sample for
this analysis.  First, galaxies were required to have secure
spectroscopic redshifts $z_{\rm spec} = 1.0 - 3.1$, allowing for
coverage of $\pa$ and/or $\pb$.  The redshift criterion reduces the
CEERS sample size from 252 unique objects with extracted spectra to
90.  Second, galaxies that may harbor AGN were excluded based on
having $\nii/\ha \ge 0.5$ or significant broad components to their
emission lines.  The exclusion of possible AGN reduces the sample size
further to 80 objects.  Third, any objects that had suspect line or
SED fits were excluded, resulting in 71 galaxies.  Finally, any
galaxies that did not have coverage of $\ha$, $\hb$, and either $\pa$
or $\pb$ were excluded.  This last criterion results in a final sample
of 63 galaxies.  

\begin{figure*}
  \epsscale{1.1}
  \plotone{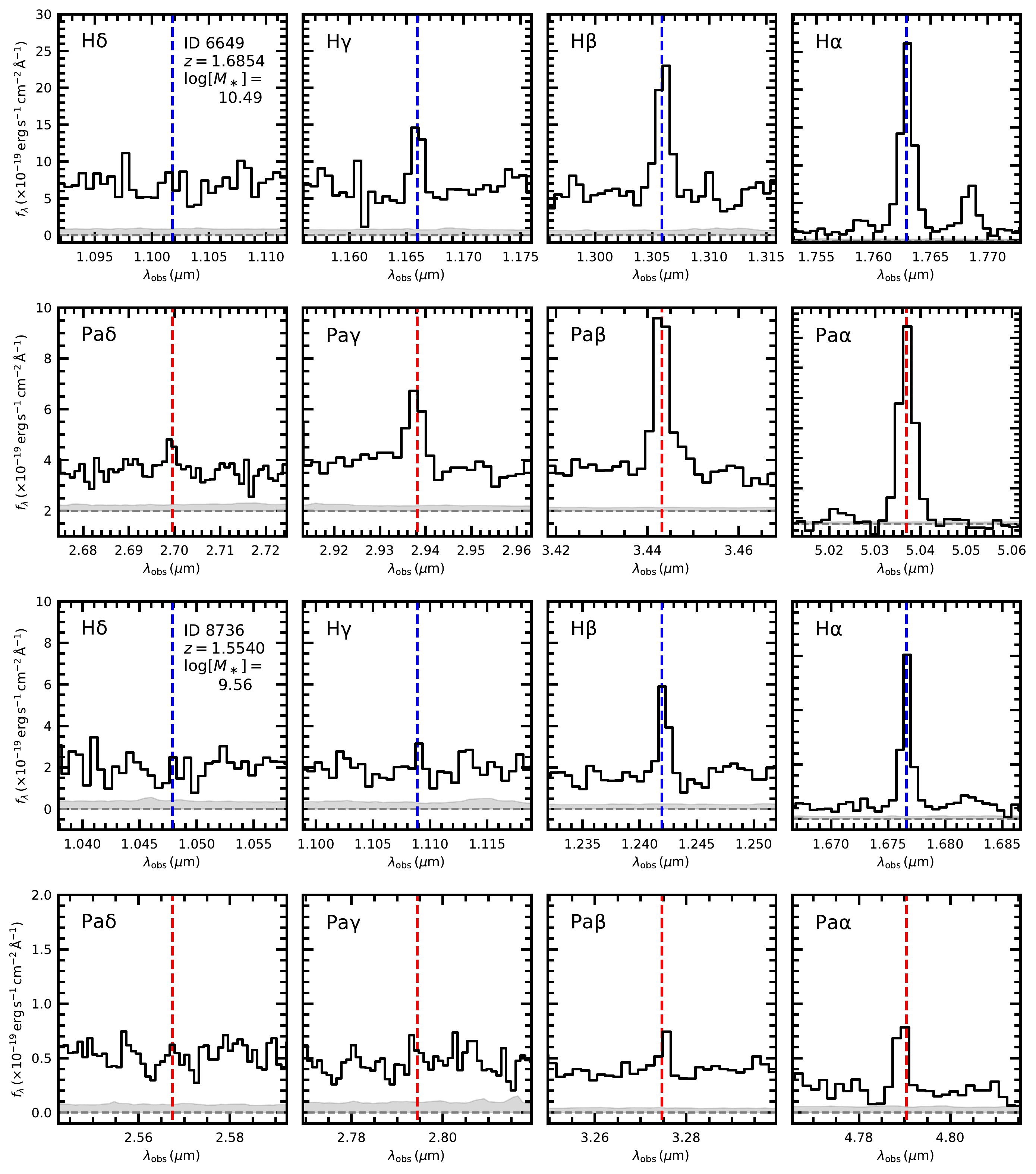}
    \caption{Example spectra of the Balmer and Paschen lines for two
      objects in the sample, CEERS ID 6649 (top two rows) and CEERS ID
      8736 (bottom two rows).  The error spectrum is shown in gray.
      The vertical scalings on the $\ha$ and $\pa$ panels have been
      increased by a factor of $2-4$ for clarity.}
    \label{fig:spec}
\end{figure*}

Of these, 61 have coverage of $\pb$, 28 have coverage of $\pa$, and 26
have coverage of both lines.  The redshift distribution of the final
sample is shown in Figure~\ref{fig:zhist}.  Of the 63 galaxies in the
sample, 38 have $\ge 3\sigma$ detections of $\ha$, $\hb$, and either
$\pa$ or $\pb$ (i.e., the ``line-detected sample'').  The bulk of the
analysis focuses on these 38 galaxies, but the entire sample of 63
galaxies is also considered when forming composite spectra.  The 63
galaxies have SED-based SFRs (Section~\ref{sec:sed}) in the range
$\approx 0.1$ to $\approx 110$\,$M_\odot$\,yr$^{-1}$, with a median of
$\simeq 2$\,$M_\odot$\,yr$^{-1}$.  The stellar masses vary in the
range $\log[M_\ast/M_\odot] = 7.03 - 10.62$, with a median of
$\log[M_\ast/M_\odot] = 9.25$.  Figure~\ref{fig:spec} shows the first
four Balmer and Paschen lines in the spectra of two galaxies in the
sample: a brighter galaxy with a higher $M_\ast$ and nebular reddening
(CEERS ID 6649), and a fainter galaxy with a lower $M_\ast$ and
nebular reddening close to zero (CEERS ID 8736).

\section{\bf PASCHEN-LINE CONSTRAINTS ON NEBULAR REDDENING AND SFRs}
\label{sec:reddening}

\subsection{The Balmer Decrement and Line Ratios Including Paschen Lines}
\label{sec:linerats}

\begin{figure*}
  \epsscale{1.1}
  \plotone{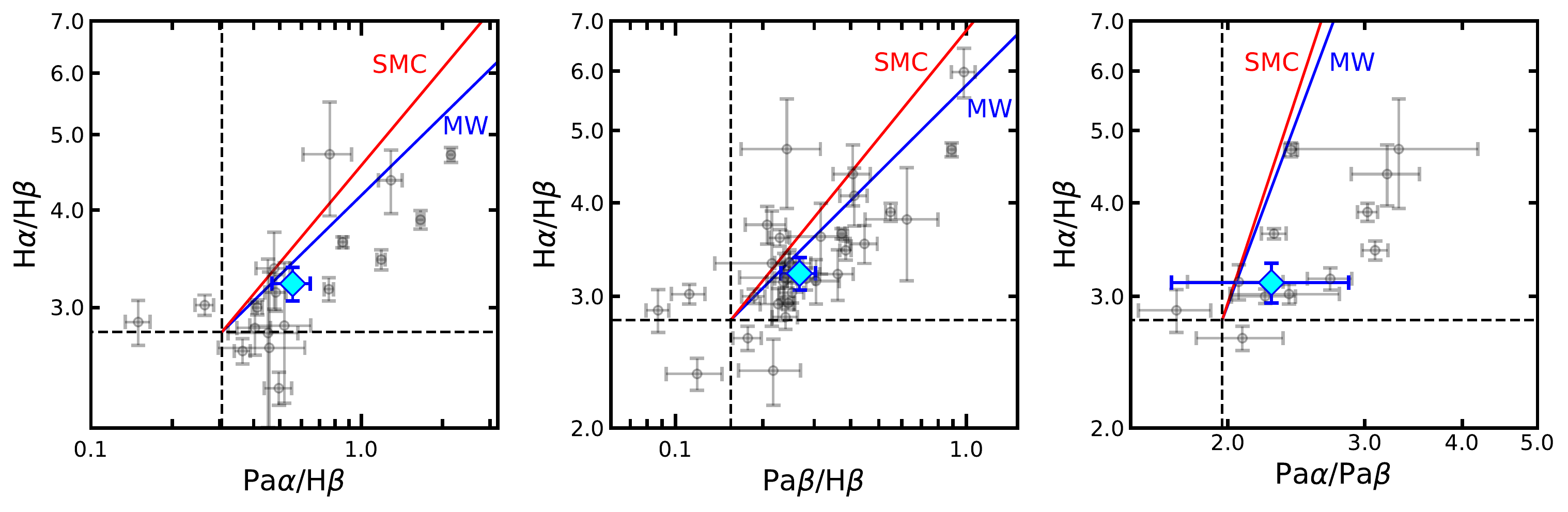}
    \caption{Relationship between the Balmer decrement, $\ha/\hb$, and
      $\pa/\hb$ (left), $\pb/\hb$ (middle), and $\pa/\pb$ (right) for
      galaxies in the line-detected sample (gray points).  The large
      diamonds indicate the average ratios measured from the composite
      spectrum of the full sample of 63 galaxies in the case of the
      left and middle panels, and the 26 galaxies with coverage of both $\pa$
      and $\pb$ in the right panel.  For comparison, the line ratios
      predicted by the SMC \citep{gordon03} and MW \citep{cardelli89}
      extinction curves are shown by the red and blue lines,
      respectively, while the dashed black lines indicate the ratios
      expected in the absence of dust attenuation: $\ha/\hb = 2.79$,
      $\pa/\hb = 0.31$, $\pb/\hb = 0.16$, and $\pa/\pb = 1.97$.}
    \label{fig:lineratios}
\end{figure*}

\begin{figure*}
  \epsscale{1.1}
  \plotone{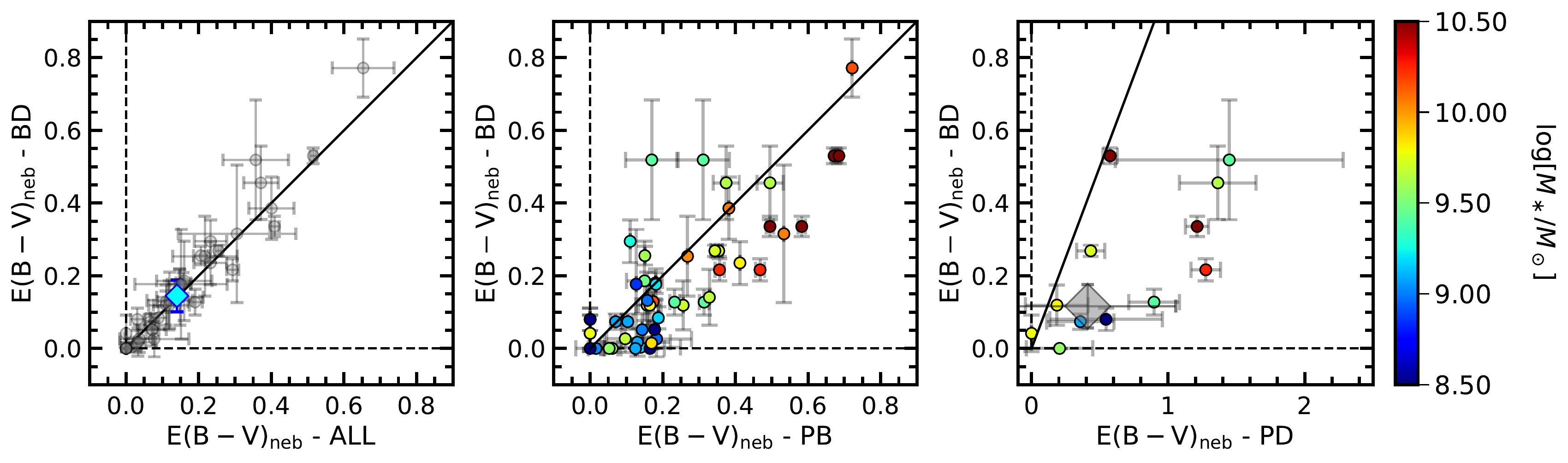}
    \caption{Comparison between the nebular reddening derived from the
      Balmer decrement and those derived from all available $\hi$
      recombination lines (left), either $\pa/\hb$ or $\pb/\hb$
      (middle), and the Paschen decrement (left).  Average reddenings
      derived from the composite spectra of the full sample of 63
      galaxies and the subsample of 26 galaxies with coverage of $\pa$
      and $\pb$ are shown by the large diamonds in the left and right
      panels, respectively.  For context, points in the middle and
      left panels are color-coded by stellar mass.  The dashed
      horizontal and vertical lines denote zero reddening, and the
      solid line indicates equality.}
    \label{fig:ebmvneb}
\end{figure*}

Given the ubiquitous use of the Balmer decrement, $\ha/\hb$, to
constrain the nebular reddening and SFRs of high-redshift galaxies, it
is useful to first assess how well it correlates with line ratios that
include the Paschen lines.  In the absence of dust, the ratios of the
different recombination lines are fixed at a given temperature and
density.  Thus, comparing the observed line ratios with the intrinsic
ones allows us to infer the reddening towards the nebular regions from
which those lines originate.  The intrinsic (dust-free) ratios assumed
in this analysis are listed in Table~\ref{tab:nebinfo}.
Figure~\ref{fig:lineratios} compares the Balmer decrements with
$\pa/\hb$ and $\pb/\hb$ for the 38 galaxies in the line-detected
sample.\footnote{There are two (of 38) galaxies in the line-detected
  sample with ratios lying $\ga 3\sigma$ below the intrinsic ratios
  expected in the absence of dust attenuation, suggesting additional
  systematic error associated with the relative flux calibration
  between gratings for these two objects.}  The right panel shows the
relationship between the Balmer decrement and the Paschen decrement,
$\pa/\pb$, for the 12 galaxies in the line-detected sample that have
$\ge 3\sigma$ detections of both Paschen lines.  These 12 galaxies are
at redshifts where $\ha$ and $\hb$ lie in the bluest grating, F100LP,
and $\pa$ and $\pb$ lie in the reddest grating, F290LP.  As such, the
ratios $\ha/\hb$ and $\pa/\pb$ for these galaxies do not suffer from
uncertainties in the relative flux calibration between gratings, and
therefore serve as a useful check of reddening inferred from ratios
consisting of lines observed in different gratings (e.g., $\pa/\hb$ or
$\pb/\hb$).

Shown in all three panels of Figure~\ref{fig:lineratios} are the
predicted line ratios as a function of reddening for the Small
Magellanic Cloud (SMC; \citealt{gordon03}) and Milky Way (MW;
\citealt{cardelli89}) extinction curves.\footnote{For the SMC curve,
  we assumed the functional form of \citet{gordon03} at $\lambda <
  1$\,$\mu$m.  At $\lambda > 1$\,$\mu$m, a $1/\lambda$ functional form
  for the curve was assumed.  The \citep{calzetti00} curve predicts
  line ratios that are similar to those of the MW.}  Spearman tests
indicate strong positive correlations between the line ratios, with
likelihoods of a null correlation of $p = 1.7\times 10^{-4}$ and
$p=7.2\times 10^{-6}$ between the Balmer decrement and $\pa/\hb$ and
$\pb/\hb$, respectively, and $p = 1.4\times 10^{-3}$ between the
Balmer and Paschen decrements.  The strength of the correlations
between the Balmer decrement and line ratios involving the Paschen
lines suggests that the former is sensitive to the overall nebular
dust attenuation of the galaxies.  In other words, the fraction of
star formation that is optically thick in the Balmer lines is not so
high as to cause a decoupling between the reddening deduced from the
Balmer lines and the reddening deduced from the less-dust-sensitive
Paschen lines.

Individually, most galaxies have line ratios statistically consistent
with the predictions of the MW extinction curve.  However, when
considering the line-detected sample as a whole, the average
$\pa/\hb$, $\pb/\hb$, and $\pa/\pb$ appear to lie systematically
higher than the predictions of either the SMC or MW curves---or the
curve derived by \citealt{reddy20} for $z\sim 2$ galaxies, which is
similar to that of the MW---at a given Balmer decrement.  On the other
hand, the average $\pa/\hb$ and $\pb/\hb$ measured from the composite
spectrum of all 63 galaxies are consistent within $3\sigma$ of the MW
predictions given the average $\ha/\hb$ of these galaxies.  The
composite spectrum of the 26 galaxies with coverage of both $\pa$ and
$\pb$ also indicates average ratios that are statistically consistent
with the MW and SMC extinction curves.  These results suggest that
some fraction of galaxies in the line-detected sample may harbor star
formation that is optically thick in the Balmer lines, a point we
return to below.

\subsection{Dependence of Nebular Reddening on Line Ratios}
\label{sec:nebredconstraints}

We next explored the extent to which the nebular reddening depends on
the specific line ratios used to derive it.  In the following
discussion, $\ebmvneb$-BD denotes the nebular reddening derived from
the Balmer decrement, $\ha/\hb$.  We also considered the reddening
derived from either $\pa/\hb$ or $\pb/\hb$, $\ebmvneb$-PB; the
reddening derived from the Paschen decrement, $\ebmvneb$-PD; and the
reddening derived from all available Balmer and Paschen lines,
$\ebmvneb$-ALL.  For the latter, we included $\ha$, $\hb$, $\hg$,
$\hd$, $\pa$, $\pb$, $\pg$, and $\pd$, and used $\chi^2$ minimization
to determine the best-fit $\ebmvneb$ that comes closest to reproducing
all the line ratios given their uncertainties.\footnote{Of the 63
  galaxies in the full sample, the numbers with $\ge 3\sigma$
  detections of $\ha$, $\hb$, $\hg$, $\hd$, $\pa$, $\pb$, $\pg$, and
  $\pd$ are 63, 53, 35, 26, 21, 38, 29, and 12, respectively.}  The MW
extinction curve was assumed for all $\ebmvneb$ calculations---the
reddenings are relatively insensitive to the choice of extinction or
attenuation curves as their shapes are very similar in the optical and
near-IR.  Uncertainties in nebular reddening are derived from
perturbing the line fluxes by their errors many times and computing
the spread in reddening derived from these many realizations.

Figure~\ref{fig:ebmvneb} shows that $\ebmvneb$-BD agrees well with
$\ebmvneb$-ALL within the uncertainties.  This result is expected
given that, relative to other lines in the fit, the high $S/N$ of
$\ha$ and $\hb$ causes these lines to dominate the fit to $\ebmvneb$.
Meanwhile, $\ebmvneb$-PB is generally consistent within $3\sigma$ of
$\ebmvneb$-BD, though there are some notable outliers for which
$\ebmvneb$-PB $>$ $\ebmvneb$-BD.  These outliers are ones for which
$\pa/\hb$ or $\pb/\hb$ significantly exceed the values predicted for
their $\ha/\hb$ (Section~\ref{sec:linerats};
Figure~\ref{fig:lineratios}).  When considering only the Paschen lines
in the $\ebmvneb$ derivation, i.e., the Paschen decrement, we find the
inferred $\ebmvneb$-PD for individual galaxies in the line-detected
sample (median value of $\ebmvneb$-PD $\simeq 0.56$) general exceed
the estimates from the Balmer decrement (median value of $\ebmvneb$-BD
$\simeq 0.17$), again suggesting the presence of some star formation
that is optically thick in the Balmer lines.  The sensitivity of the
Paschen lines to regions that may be more attenuated (relative to the
Balmer lines) has also been observed in nearby galaxies with $\ha$ and
$\pb$ line maps \citep{gimenez22}.  The average $\ebmvneb$-PD derived
from the composite spectrum of the 26 galaxies with coverage of both
Paschen lines lies within $3\sigma$ of $\ebmvneb$-BD.  The similar or
higher nebular reddening derived from the Paschen lines versus the
Balmer lines implies differences in nebular and stellar reddening at
least as large as those reported in previous studies that were based
on Balmer decrements (e.g., \citealt{calzetti97, calzetti00,
  forster09, yoshikawa10, wuyts11, wild11, kreckel13, kashino13,
  price14, reddy15, debarros16, buat18, reddy20, shivaei20a}).

\subsection{SFRs from Balmer and Paschen Lines}
\label{sec:sfrnebcompare}

As the Paschen lines are less affected by dust, they are more directly
connected to the bolometric SFRs of galaxies compared to the Balmer
lines.  Paschen-line SFRs were computed by dust correcting either the
$\pa$ or $\pb$ luminosities according to $\ebmvneb$-PB, and converting
the dust-corrected luminosities to SFRs using the following equations: 
\begin{equation}
{\rm SFR(\pa)} [M_\odot\,{\rm yr}^{-1}] = C(\pa) \times L(\pa) [{\rm erg}\,{\rm s}^{-1}]
\label{eq:sfrpa}
\end{equation}
and 
\begin{equation}
{\rm SFR(\pb)} [M_\odot\,{\rm yr}^{-1}] = C(\pb) \times L(\pb) [{\rm erg}\,{\rm s}^{-1}],
\label{eq:sfrpb}
\end{equation}
where $C(\pa) = 1.95\times 10^{-41}$ ($3.90\times 10^{-41}$) and
$C(\pb) = 3.84\times 10^{-41}$ ($7.67\times 10^{-41}$) for a $Z_\ast =
0.001$ ($0.02$) Binary Population and Stellar Synthesis
(BPASS) v2.2.1 model \citep{eldridge17, stanway18} with an upper mass
cutoff of the IMF of $100$\,$M_\odot$ \citep{reddy22}.  Following
\citet{du18}, we adopted the $Z_\ast = 0.02$ (solar metallicity)
conversion factor for all galaxies at $z\le 1.4$, galaxies at
$1.4<z\le 2.7$ with $M_\ast > 10.45$, and galaxies at $z>2.7$ with
$M_\ast > 10.66$.  The $Z_\ast = 0.001$ (subsolar metallicity)
conversion factor was adopted for all other galaxies.  The same
redshift and mass criteria were used to compute $\ha$-based SFRs,
where the $\ha$ luminosity was corrected for dust using $\ebmvneb$-BD,
and assuming
\begin{equation}
{\rm SFR(\ha)} [M_\odot\,{\rm yr}^{-1}] = C(\ha) \times L(\ha) [{\rm erg}\,{\rm s}^{-1}],
\end{equation}
where $C(\ha) = 2.14 \times 10^{-42}$ ($4.27\times 10^{-42}$) for the
$Z_\ast = 0.001$ ($0.02$) models discussed above.  Direct
comparisons between Balmer-based and Paschen-based SFRs are unaffected
by the stellar-metallicity assumptions as both sets of SFRs will be
affected in an identical manner.  The errors in SFRs include the
uncertainties in line fluxes and reddening.

For the 12 galaxies with measured
Paschen decrements, the SFRs computed by dust correcting the $\pa$
luminosity according to $\ebmvneb$-PD lie within $\approx 15\%$ of
those computed using $\ebmvneb$-PB.  Specifically, while in some cases
$\ebmvneb$-PD may be significantly redder than $\ebmvneb$-PB for a
given galaxy, the attenuation in magnitudes at the wavelength of
$\pa$, $\lambda = 1.88$\,$\mu$m, is typically $\la 0.25$\,mag (with
the maximum $A(\pa) \simeq 0.7$\,mag for the reddest galaxy in the
sample), resulting in small differences in the resulting SFRs.  We
return to this point below.  

\begin{figure}
  \epsscale{1.0}
  \plotone{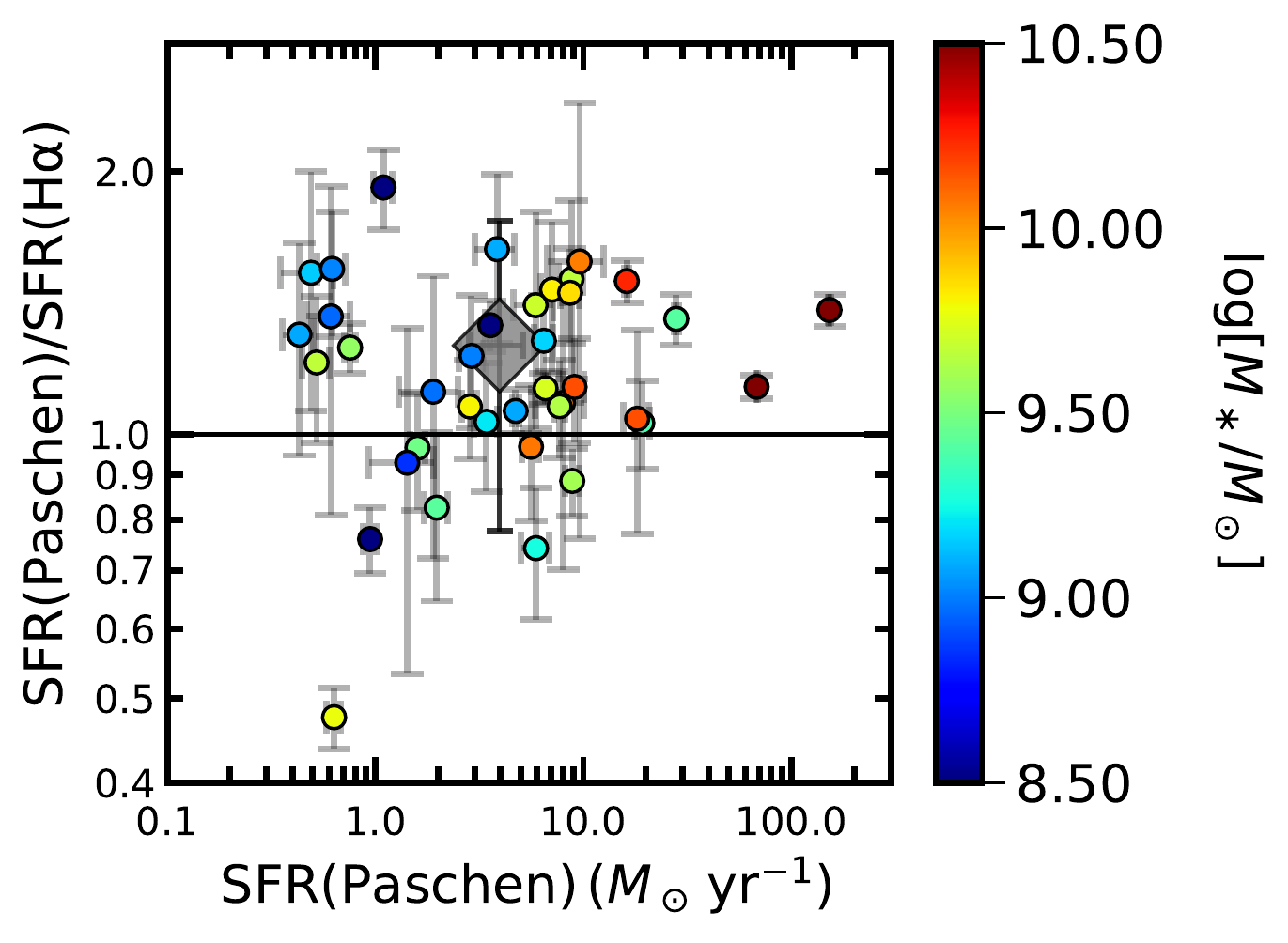}
    \caption{Distribution of the ratio of Paschen-line SFRs to $\ha$
      SFRs for the 38 galaxies in the line-detected sample,
      color-coded by stellar mass.  The large diamond indicates the
      average values computed from the composite spectrum of the 63
      galaxies in the full sample.  The horizontal line indicates
      equality between the Paschen-line and $\ha$ SFRs.}
    \label{fig:sfrneb}
\end{figure}

The ratios of the Paschen-line SFRs to $\ha$ SFRs are shown in
Figure~\ref{fig:sfrneb}.  The average ratio derived from the composite
spectrum of the full sample is also shown in the figure.  The two
galaxies with the highest SFRs and $M_\ast$ in the sample have
Paschen-line SFRs that clearly exceed their $\ha$ SFRs, and
$\ebmvneb$-PB that are significantly redder than $\ebmvneb$-BD
(Figure~\ref{fig:ebmvneb}).  These results suggest the presence of
star formation that is optically thick in the Balmer lines.  Not
surprisingly, the fainter lines characteristic of the lower luminosity
and lower $M_\ast$ galaxies yield uncertainties in the SFR ratios that
are too large to draw a robust conclusion regarding the fraction of
optically-thick star formation in these galaxies.  Based on
measurements from the composite spectrum, the average Paschen-line SFR
is $\approx 25\%$ larger than the average $\ha$ SFR of galaxies in the
sample, but all galaxies have Paschen-line SFRs that are within a
factor of 2 of the $\ha$ SFRs.  Deeper spectroscopy with higher $S/N$
measurements of the Paschen lines---and the continuum, which aids in
accurate flux calibration---is needed to robustly constrain the
fraction of optically-thick star formation in these galaxies.

\begin{figure}
  \epsscale{1.0}
  \plotone{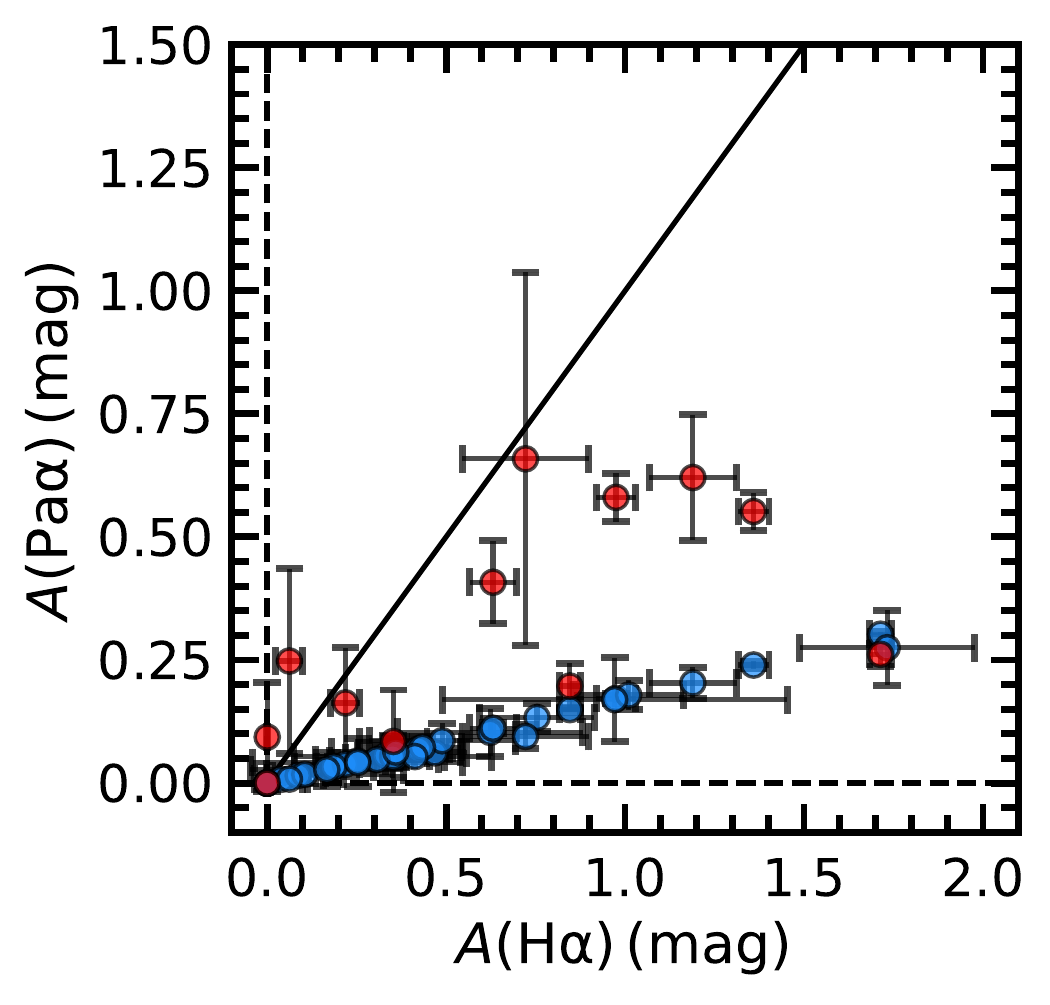}
    \caption{Comparison of the attenuation in magnitudes of $\ha$,
      assuming $\ebmvneb$-BD, and $\pa$, assuming $\ebmvneb$-PB (blue
      points).  The red points reflect the distribution of attenuation
      of $\pa$ if we assume $\ebmvneb$-PD.  }
    \label{fig:attmag}
\end{figure}

An expected yet important issue to emphasize is the sensitivity of the
Paschen lines to bolometric SFR.  This aspect is illustrated in
Figure~\ref{fig:attmag}, where the attenuation in magnitudes of $\ha$
(assuming $\ebmvneb$-BD) and $\pa$ (assuming $\ebmvneb$-PB) are shown
for the line-detected sample.  The typical attenuation of $\pa$ is
only $\sim 0.1$\,mag, while it is closer to $\sim 0.5$\,mag for $\ha$.
Even with the more extreme values of reddening derived from the
Paschen decrement, $\ebmvneb$-PD, for the 12 galaxies with significant
detections of $\pa$ and $\pb$, the derived attenuation of $\pa$ is
still smaller than that of $\ha$, and even more so had we assumed
$\ebmvneb$-PD in computing the attenuation of $\ha$.  Clearly, the
Paschen lines, and in particular $\pa$, allow for a more direct probe
of bolometric SFR that is less sensitive to dust reddening, an aspect
that we take advantage of in the next section to probe the
relationship between dust obscuration and UV reddening.

\section{\bf PASCHEN-INFERRED DUST OBSCURATION AND THE UV CONTINUUM SLOPE}
\label{sec:irx}

\begin{figure}
  \epsscale{1.2}
  \plotone{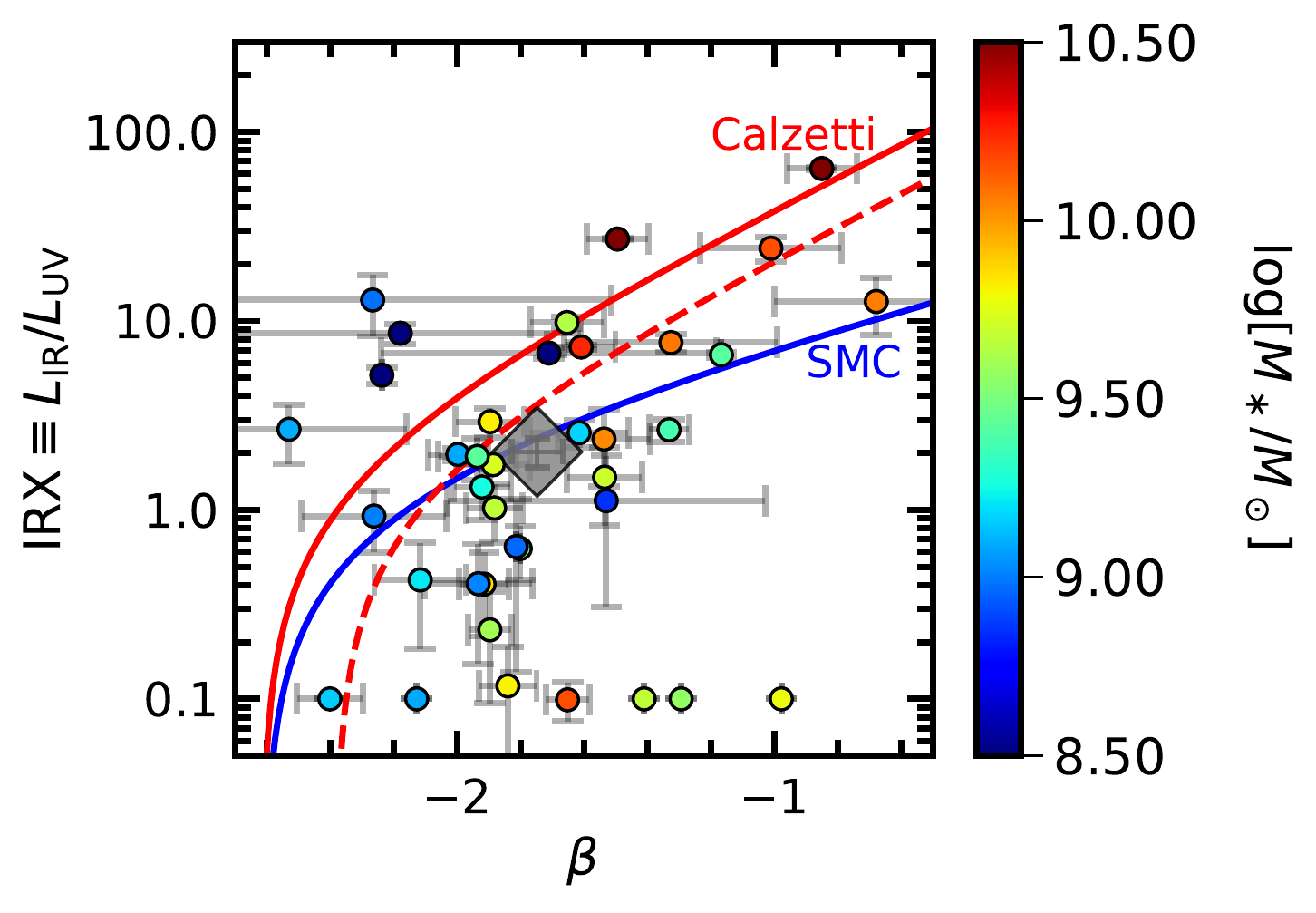}
    \caption{The IRX-$\beta$ diagram for the 38 galaxies in the
      line-detected sample.  Shown are the predicted relationships
      between IRX and $\beta$ for the \citet{calzetti00} attenuation
      curve and the SMC extinction curve as applied to a $Z_\ast
      \simeq 0.001$ stellar population \citep{reddy18a}.  The dashed
      red curve shows the predicted relation between IRX and $\beta$
      for the \citet{calzetti00} attenuation curve as applied to a
      $Z_\ast \simeq 0.02$ stellar population (the stellar metallicity
      affects the intrinsic UV slope, $\beta_0$).  Data points are
      color coded by stellar mass.  The large diamond shows the
      average value inferred from the composite spectrum of the 63
      galaxies in the full sample.}
    \label{fig:irx}
\end{figure}

The spectral slope of the far- to near-UV stellar continuum, $\beta$,
where $f_{\lambda}\propto \lambda^\beta$, has long been used as a
proxy for dust obscuration (e.g., \citealt{calzetti94, meurer99,
  reddy06b, daddi07a, overzier11, reddy12a, buat12, bouwens16b,
  reddy18a, fudamoto20}).  The sensitivity of the Paschen lines to
bolometric SFR offers a new opportunity to reevaluate the relationship
between $\beta$ and dust obscuration.  The best-fit SED of each galaxy
(Section~\ref{sec:sed}) was used to calculate the flux density at
$1600$\,\AA, which in turn was converted to an (unobscured) UV
luminosity.  This unobscured UV luminosity was converted to a UV SFR
based on the same redshift and mass criteria discussed in Section~\ref{sec:sfrnebcompare},
where we assumed
\begin{equation}
{\rm SFR(UV)} [M_\odot\,{\rm yr}^{-1}] = C({\rm UV}) \times \nu L_\nu [{\rm erg}\,{\rm s}^{-1}],
\label{eq:sfruv}
\end{equation}
and $C({\rm UV}) = 3.72\times 10^{-44}$ ($4.17\times 10^{-44}$) for
the $Z_\ast = 0.001$ ($0.02$) BPASS models with an age of $100$\,Myr,
typical of the ages derived for galaxies in the sample.  This UV SFR
was subtracted from the (dust-corrected) Paschen-line SFR
(Section~\ref{sec:sfrnebcompare}) to yield an obscured SFR.  This
obscured SFR was then converted to infrared (IR) luminosity using the
relation from \citet{kennicutt98}, modified for a \citet{chabrier03}
IMF.  There are 5 of 38 galaxies for which the UV SFR exceeds the
Paschen-line SFR.  In these cases, the IR luminosity was fixed to
$10\%$ of the unobscured UV luminosity.  Thus, for each galaxy, we
have an estimate of the ratio of the IR and UV luminosities, i.e.,
$\lir/\luv$, or IRX.  The photometry corresponding to the rest-frame
UV was used to compute $\beta$ for each galaxy.  Uncertainties were
computed by perturbing the photometric fluxes many times and
recomputing $\beta$.

The relationship between dust obscuration, parameterized by IRX, and
$\beta$, i.e., the IRX-$\beta$ relation, for galaxies in the
line-detected sample is shown in Figure~\ref{fig:irx}, along with the
predicted relations assuming the \citet{calzetti00} and SMC curves and
an intrinsic $\beta_0 = -2.62$ appropriate for the $Z_\ast = 0.001$
BPASS model discussed above.  While there is relatively large spread
in IRX at a given $\beta$---which likely correlates with gas-phase
metallicity and/or mass (e.g., \citealt{reddy06b, buat12, zeimann15,
  bouwens16a, reddy18a, shivaei20b})---the data points generally
scatter around the IRX-$\beta$ relation predicted for a subsolar
stellar metallicity model and the SMC extinction curve (see
\citealt{reddy18a}).  The average IRX and $\beta$ for the full sample
of 63 galaxies (large diamond in Figure~\ref{fig:irx}) are also
consistent with the predictions of the SMC curve.  These results
confirm previous studies of stacked {\em Herschel} far-IR data
suggesting that the reddening of the stellar continuum of $z\sim 2$
galaxies with subsolar stellar metallicities is in accordance, on
average, with the SMC extinction curve \citep{reddy18a}.  We also find
that the most massive galaxies in our sample ($M_\ast \ga
10^{10}$\,$M_\odot$) exhibit redder UV slopes and higher IRX that may
be more consistent with a \citet{calzetti00} curve and a solar
metallicity model (shown by the dashed red line in
Figure~\ref{fig:irx}), confirming previous investigations that found
similar results at high redshift (e.g., \citealt{reddy10,
  bouwens16a}).

\begin{figure}
  \epsscale{1.2}
  \plotone{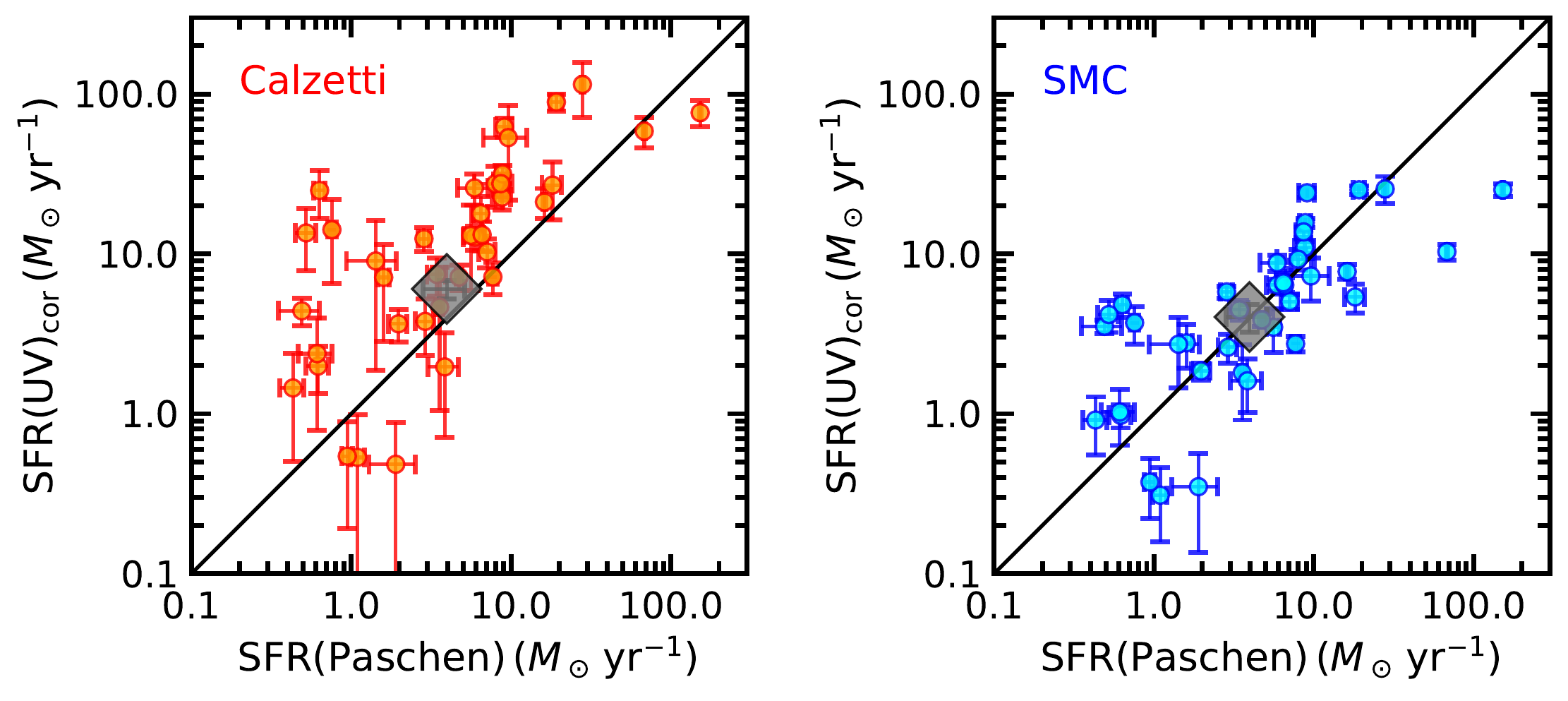}
    \caption{Dust-corrected UV SFRs versus Paschen-line SFRs, where
      the former have been corrected for dust assuming the
      \citet{calzetti00} attenuation curve and a solar metallicity
      stellar population (left), and the SMC extinction curve and a
      subsolar metallicity stellar population (see
      Section~\ref{sec:sed}).  The average SFRs for the entire sample
      of 63 galaxies are shown by the large diamonds.}
    \label{fig:sfrcompare}
\end{figure}

These results are also evident in a direct comparison of the
Paschen-based SFRs and the UV based SFRs, where the latter have been
corrected for dust using either the \citet{calzetti00} dust
attenuation curve with a solar metallicity model or the SMC extinction
curve with a subsolar metallicity model (Figure~\ref{fig:sfrcompare}).
The average Paschen-line SFR computed from the composite spectrum of
all 63 galaxies is in good agreement with the average dust-corrected
UV SFR in both cases.  However, for the line-detected sample, the
median residual of the SFRs relative to the line of equality is
significantly reduced when assuming the SMC extinction curve (median
residual of $-0.05$) versus the \citet{calzetti00} curve (median
residual of $1.67$).  The SFR comparison suggests the general
applicability of the SMC extinction curve for the UV continuum
reddening, but note that there are some cases (e.g., such as the two
galaxies with the highest Paschen-line SFRs in the sample) where the
\citet{calzetti00} curve may provide a better description.  

Finally, note the apparent increase in scatter between UV- and
Paschen-based SFRs for galaxies at the lower end of the SFR
distribution (i.e., Paschen-based SFRs $\la 2$\,$M_\odot$\,yr$^{-1}$),
irrespective of the attenuation curve used to correct the former.
This effect has also been seen in comparisons between UV and
Balmer-line SFRs, as well as in low-redshift ($z\la 0.3$) galaxies
with UV and $\pb$ measurements \citep{cleri22}, and has commonly been
attributed to stochastic variations in SFR for faint and low-mass
galaxies at high redshift (e.g., \citealt{dominguez15, guo16, emami19,
  atek22}).  Our results are the first at $z>1$ to show that this
increase in scatter between UV and Balmer-line SFRs persists when
using the less-dust-sensitive Paschen lines to infer the SFR.

\section{\bf CONCLUSIONS}
\label{sec:conclusions}

We present the first analysis of the Paschen lines for a substantial
sample of 63 galaxies at $z=1.0-3.1$ observed with {\em JWST}/NIRSpec
as part of the CEERS survey.  Using a combination of Balmer and
Paschen lines, we constrain the dust reddening and SFRs of these
galaxies.  The Balmer decrement, $\ha/\hb$, is strongly correlated
with ratios involving the Paschen lines, including $\pa/\hb$,
$\pb/\hb$, and the Paschen decrement, $\pa/\pb$, confirming that the
Balmer decrement is sensitive to the overall reddening within these
galaxies.  The measured Paschen-line ratios suggest nebular reddening
that is typically larger than that obtained with the Balmer decrement,
at least for galaxies with higher SFRs and stellar masses.
Consequently, we find marginal evidence that Paschen-based SFRs are
$\approx 25\%$ larger than those deduced from the Balmer lines alone
for such galaxies, though deeper observations are needed to confirm
this.  Using the Paschen-based SFRs, we investigate the relationship
between dust obscuration, parameterized by the IRX ratio, and UV
continuum slope, $\beta$.  These measurements, along with direct
comparisons of Paschen-line and dust-corrected UV SFRs, confirm
previous studies that have found that the SMC extinction curve
provides an adequate description of the reddening of the UV continuum
in subsolar-metallicity high-redshift galaxies, at least on average.

Larger and more representative samples of galaxies with Paschen-line
observations will enable us to extend this analysis to directly
constrain the shape of the nebular attenuation curve at high redshift
\citep{reddy20, rezaee21, prescott22}, and nebular reddening and SFRs
and their relationship to stellar continuum reddening.  The
sensitivity of the Paschen lines to bolometric SFR makes them a
critical component for calibrating infrared-based SFRs with {\em
  JWST}/MIRI observations; investigating stochasticity in SFR,
particularly for low-mass galaxies; and providing stringent
constraints on dust-corrected line ratios that are commonly used to
infer the physical state of the ISM, including the ionization
parameter and gas-phase metallicity.

\acknowledgements

This work is based on observations made with the NASA/ESA/CSA James
Webb Space Telescope.  The data were obtained from the Mikulski
Archive for Space Telescopes at the Space Telescope Science Institute,
which is operated by the Association of Universities for Research in
Astronomy, Inc., under NASA contract NAS5-03127 for {\em JWST}.
Support for this work was also provided through the NASA Hubble
Fellowship grant HST-HF2-51469.001-A awarded by the Space Telescope
Science Institute, which is operated by the Association of
Universities for Research in Astronomy, Incorporated, under NASA
contract NAS5-26555.  The Cosmic Dawn Center is funded by the Danish
National Research Foundation (DNRF) under grant \#140.  Cloud-based
data processing and file storage for this work is provided by the AWS
Cloud Credits for Research program.

%\bibliographystyle{aasjournal}
%\bibliography{apj-jour,myrefs}

\end{document}